\documentclass[12pt]{article}
\usepackage{amsmath,amssymb}
\usepackage{bm}
\usepackage{color}
\usepackage{cite}
\usepackage{mathtools}

\usepackage{multirow}

\def\Slash#1{\rlap/#1}
\def\Slash#1{\rlap{\kern .1em /}#1}

\usepackage{here}
\usepackage {cancel}

\begin{document}

	\title{
	\begin{flushright}
		\ \\*[-80pt]
		\begin{minipage}{0.2\linewidth}
			\normalsize
		\end{minipage}
	\end{flushright}
	%
	{\Large \bf  Prediction of the CP Phase $\delta_{CP}$ in the Neutrino Oscillation and an Axion-less Solution to the Strong CP Problem
		\\*[20pt]
	}
}

\author{Morimitsu Tanimoto $^{1}$
	\footnote{email: tanimoto@muse.sc.niigata-u.ac.jp}
	\hskip 0.5 cm 
 and Tsutomu T. Yanagida $^{2,3}$
 \footnote{email: tsutomu.tyanagida@sjtu.edu.cn
 }
	\\*[20pt]
	\centerline{
		\begin{minipage}{\linewidth}
			\begin{center}
					$^1${\it \normalsize
					Department of Physics, Niigata University, Niigata 950-2181, Japan }
				\\*[10pt]
			\end{center}
      \begin{center}
					$^2${\it \normalsize
					Kavli Institute for the Physics and Mathematics of Universe (WPI),
     \\*[10pt] University of Tokyo, Kashiwa 277-8583, Japan }
				\\*[10pt]
			\end{center}
    \begin{center}
					$^3${\it \normalsize
	Tsung-Dao Lee Institute and School of Physics and Astronomy,
     \\*[10pt] 
      Shanghai Jiao Tong University, China}
				\\*[10pt]
			\end{center}
	\end{minipage}}
	\\*[30pt]}
\date{\today\\	\vskip 1 cm
{\small \bf Abstract}
	\begin{minipage}{0.9\linewidth}
		\medskip
		\medskip
		\small
	A model of the axion-less solution to the strong CP problem has been recently constructed based on the $\bf T^2/\mathbb{Z}_3$ orbifold in the six dimensional space-time. We extend, in this paper, the model to include the lepton sector with three right-handed neutrinos. A crucial point in the extended model is that we have only one CP-violating phase and thus the CP phase in the neutrino oscillation can be predicted by the CP phase in the CKM matrix. The CP phase $\delta_{CP}$ in the PMNS matrix is predicted as  {$\delta_{CP}\simeq 192^\circ - 195^\circ$}
   which can be tested in the near future experiments. We also discuss the effective mass $m_{\beta\beta}$ responsible for the neutrino-less double beta decay of nucleus, which is predicted as $m_{\beta\beta} \simeq 8-11$\,meV.
   {We consider the leptogenesis and confirm that our model generates the right sign of the baryon asymmetry if the first family right-handed neutrino is the lightest among the heavy right-handed neutrinos.}
	\end{minipage}}

\begin{titlepage}
\maketitle
\thispagestyle{empty}
\end{titlepage}


\newpage

\maketitle

\section{Introduction}

The cosmological baryon asymmetry is one of the most fundamental parameters in the Universe. However, the absolute value and the sign of the baryon asymmetry depends on the details of the masses of heavy particles, unknown coupling to the light particles and the reheating temperature $T_R$ in the early universe, for example. If the CP violation at high energies is interrelated directly with the CP violation in low energies, the presence of the cosmological baryon asymmetry strongly predicts the presence of CP violation in the low-energy processes such as the K-meson decay, the neutrino oscillation and etc. However, most of the baryogenesis mechanisms contain many independent CP-violating phases and hence the CP violation for the cosmological baryogenesis is, unfortunately, independent of the low-energy CP-violating phases. Therefore, the presence of the baryon asymmetry does not explain the existence of the CP violation in the K-meson decay and does not assures us a non-vanishing CP violation in neutrino oscillation experiments.

Recently, Liang, Okabe and one of the authors (T.T.Y.) have proposed a simple model for an axion-less solution to the strong CP problem
\cite{Liang:2024wbb}, motivated by a consistent three zeros texture of the quark mass matrix  in \cite{Tanimoto:2016rqy}. In this paper, we extend the original LOY model to the lepton sector including three right-handed neutrinos $N_{i}~(i$=1-3) and consider the leptogenesis mechanism for generating the  baryon-number asymmetry\cite{Fukugita:1986hr} in the Universe. A surprising result in this extension is that we have only one CP-violating phase $\phi$  and  the cosmological CP-violating phase is directly related to  the CP
phase $\delta^{\rm CKM}_{CP}$ in the CKM matrix \cite{ParticleDataGroup:2024cfk}. Furthermore, we also show that the CP phase $\delta_{CP}$ in the neutrino oscillation is predicted by using the CP phase in the CKM matrix, which will be tested in near future experiments 
\cite{Hyper-KamiokandeWorkingGroup:2014czz}.

In section 2, we discuss the extension of the LOY model to the lepton sector.
The charged lepton mass matrix, the Dirac  neutrino mass matrix and the heavy right-handed neutrino (RHN) mass matrix are given in this section. In section 3, we examine the  neutrino mass matrix, $M_\nu$, in details. By using the measured parameters and constraints in the neutrino oscillation data, we derive all matrix elements in the $M_\nu$ and give our predictions for the neutrino sector. 
In section 4, we consider the leptogenesis mechanism for generating the cosmological baryon asymmetry and show the basic CP-violating phase $\phi$ for the leptogenesis is exactly the same as the CP phase in the CKM matrix.
The section 5 is devoted to the conclusion.


\section{Extension of the LOY model to the lepton sector}

We first briefly review the original LOY model \cite{Liang:2024wbb}. The fundamental high-energy theory is assumed to preserve the CP invariance and the original QCD vacuum angle $\theta$ is completely vanishing. However, the CP invariance is spontaneously broken down at some intermediate-energy scale, generating the CKM CP phase. The model is based on a six dimensional spacetime in which the extra two dimensions are assumed to be compactified on  $\bf T^2/\mathbb{Z}_3$ orbifold. We have three fixed points
I, I\hskip -0.5mm I and   I\hskip -0.5mm I\hskip -0.5mm I in the extra dimensions. Three families of {\bf 10}s are localized at each fixed points, separately. The first family ${\bf 5^*}$ is localized on the fixed point I\hskip -0.5mm I and other $ {\bf 5^*}$s are living in the extra two dimensional bulk. We impose a $\mathbb{Z}_2$ symmetry so that we have more than three zeros in the mass matrix for the down-type quarks.

Now, we introduce a CP and $\mathbb{Z}_2$ violating complex field $\eta$ which is assumed to be localized at the fixed point I\hskip -0.5mm I. We should stress here that the CP-violating phase $\phi$ in the quark and lepton mass matrices is given by an interaction of a new heavy doublet boson $H_{\rm I\hskip -0.5mm I}$, that is, $H_{\rm I\hskip -0.5mm I} H^\dagger(\alpha\eta + \beta \eta^*)$, where the $H$ is the standard model Higgs boson and the $\alpha$ and $\beta$ are dimension one real parameters. The heavy boson 
$H_{\rm I\hskip -0.5mm I}$ couples to the (2,\,2) element in the mass matrices we discuss below and hence the CP-violating phase is universal for all relevant matrices. This is the reason why we can connect the CP violation at low energies to the CP violation in the leptogenesis at high energies. See \cite{Liang:2024wbb} for details.
The $\mathbb{Z}_2$ charges and locations in the extra two dimensions are summarized 
for each particles in Table \ref{list}.

\begin{table}[hbtp]
\begin{center}
\begin{tabular}{|c||c|c|c|c|c|c|c|c|c|c|c|c|c|}
\hline 
	\rule[14pt]{0pt}{3pt}  
&$\bf 10_1$ &$\bf 10_2$ &$\bf 10_3$ & $\bf 5_1^*$ &$\bf 5_2^*$ & $\bf 5_3^*$ & $N_{1}$ &
$N_{2}$& $N_{3}$& $H$ & $H_{ \rm I\hskip -0.5mm I }$ & $\eta$  &$\delta$\\
\hline 
	\rule[14pt]{0pt}{3pt}  
$ \mathbb{Z}_2$&  - & + & + & + & - & + & - & + & + & + & - & - & -\\
\hline 
	\rule[14pt]{0pt}{3pt}  
location&  I & I\hskip -0.5mm I &  I\hskip -0.5mm I\hskip -0.5mm I 
&I\hskip -0.5mm I & bulk & bulk & I & I\hskip -0.5mm I &  I\hskip -0.5mm I\hskip -0.5mm I &  bulk &   I\hskip -0.5mm I  & I\hskip -0.5mm I &  I\hskip -0.5mm I\hskip -0.5mm I \\
\hline
\end{tabular}
\caption{List of $\mathbb{Z}_2$  charges and  locations  in the extra two dimensions
(fixed points I, I\hskip -0.5mm I, I\hskip -0.5mm I\hskip -0.5mm I  and bulk) for each particles.}
\label{list}
\end{center}
\end{table}

A $\mathbb{Z}_2$ odd and real field $\delta$ is introduced 
at the fixed point  I\hskip -0.5mm I\hskip -0.5mm I  so that we can reproduced the mass matrix $M_d^{(1)}$ in the left-right convention (LR) in \cite{Tanimoto:2016rqy},
\begin{align}
M_d^{(1)}=
\begin{pmatrix}
0 & a & 0 \\
a' & b \ e^{-i\phi}& c\\
0 & c' & d
\end{pmatrix} \ , \quad
\label{Md1}
\end{align}
where $a, a', b, c, c'$ and $d$ are real parameters, and $\phi$ is the CP-violating phase. Those seven parameters are determined by the measured CKM matrix and masses of the down-type quarks, which are summarized in Table \ref{tab1}, where parameters are taken to be positive without loss of generality. All mass parameters are defined at the weak scale.
\begin{table}[hbtp]
\begin{center}
\begin{tabular}{|c|c|c|c|c|c|c|}
\hline 
	\rule[14pt]{0pt}{3pt}  
 $a$ [MeV]& $a'$ [MeV] & $b$ [MeV] &  $c$ [MeV]& $c'$ [GeV]& $d$ [GeV] &$\phi$ [$\circ$]\\
\hline 
	\rule[14pt]{0pt}{3pt}  
 $15$\,-\,$17.5$ &$10$\,-\,$15$&$92$\,-\,$104$ & $78$\,-\,$95$ &$1.65$\,-\,$2.0$ 
 & $2.0$\,-\,$2.3$ &$37$\,-\,$48$\\
\hline
\end{tabular}
\caption{The allowed regions of parameters in  $M_d^{(1)}$.}
\label{tab1}
\end{center}
\end{table}

Since the determinant of $M_d^{(1)}$ is real, the physical QCD angle ${\bar \theta}$ remains at zero and hence there is no strong CP problem. A crucial point here is that all matrix elements including the CP-violating phase $\phi$ are determined very precisely from the experimental data on the CKM matrix and the down-type quark masses 
(see \cite{Tanimoto:2016rqy} for details).

\begin{figure}[h!]
\begin{minipage}[]{0.45\linewidth}
\includegraphics[width=8cm]{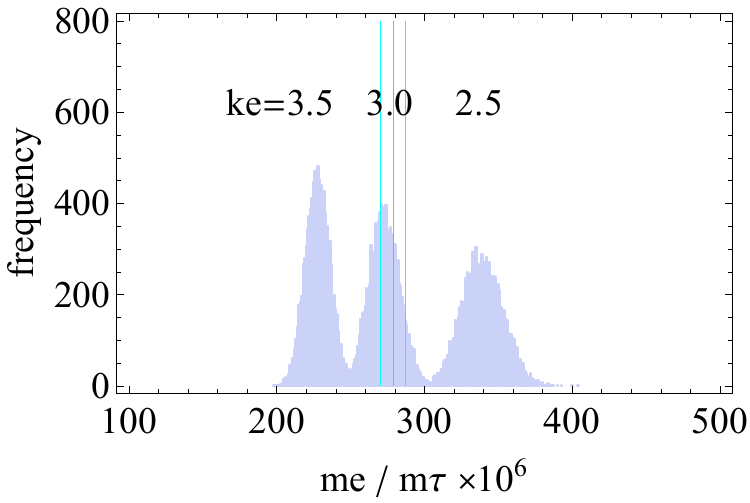}
\caption{Frequency plot versus the mass ratio $m_e/m_\tau$ for $k_e=2.5,\, 3.0,\, 3.5$.
The red and cyan lines denote the central value 
and $3\%$ error-bar of the  observed one, respectively.}
\label{fig:m1m3}
\end{minipage}
\hspace{10mm}
\begin{minipage}[]{0.45\linewidth}
\includegraphics[width=8cm]{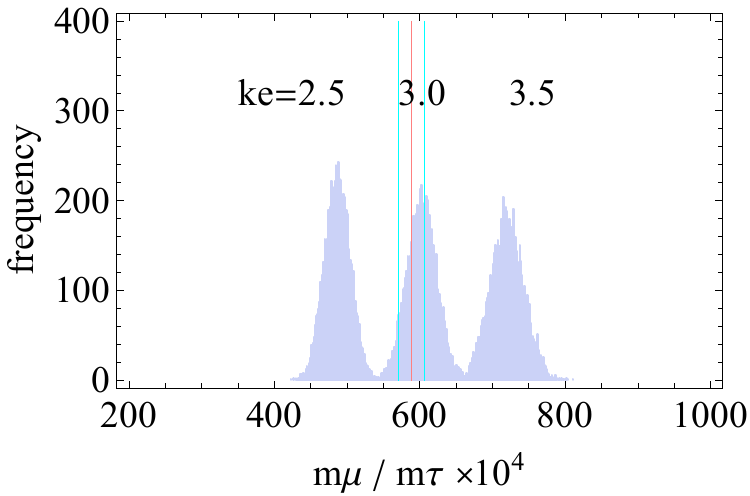}
\caption{Frequency plot versus the mass ratio $m_\mu/m_\tau$ for $k_e=2.5,\, 3.0,\, 3.5$.
The red and cyan lines denote the central value 
and $3\%$ error-bar of the  observed one, respectively.}
\label{fig:m2m3}
\end{minipage}
\end{figure}

In the above model construction we use the notation of $SU(5)$, but it does not mean the grand unification of all gauge interactions. We only gauge the standard $SU(3)\times SU(2)\times U(1)$ subgroup. However, as long as all quarks and leptons belong to the $\bf 10$ and $\bf 5^*$ we can derive the mass matrix for the charged leptons $M_e$ from $M_d$. We see easily that it is given by the transposed matrix of $M_d^{(1)}$ unless there is no large deviation from the $SU(5)$. However, if it is exactly so, the electron mass becomes too heavy. We vary the magnitude of the (2,\,2) element so that we can obtain the correct electron and muon masses. This might be reasonable since the (2,\,2) element is given by the Yukawa coupling of the heavy doublet boson $H_{\rm I\hskip -0.5mm I}$ as shown in \cite{Liang:2024wbb}.
Therefore, the charged lepton mass matrix $M_e$ is given by:
\begin{align}
M_e=
\begin{pmatrix}
0 & a' & 0 \\
a & k_e\, b \ e^{-i\phi}& c'\\
0 & c & d
\end{pmatrix}\ , \quad
\label{Me}
\end{align}
where
we multiply the (2,\,2) element by a factor $k_e$. 
{Notice that the phase factor $e^{-i\phi}$ is the same as that in $M_d$, since the factor is given by the coupling among $H_{\rm I\hskip -0.5mm I}^\dagger$, $H$ and $\alpha\eta + \beta\eta^\dagger$ as shown in \cite{Liang:2024wbb}. Here, $\alpha$ and $\beta$ are real constants.}
The result on the electron and muon masses is presented, respectively  in Figs.\,\ref{fig:m1m3} and \ref{fig:m2m3},
where the mass ratios of the electron and the muon to the tau
are plotted for $k_e=2.5,\, 3.0,\, 3.5$.\footnote{{Frequency plots in Figs.\,\ref{fig:m1m3} and \ref{fig:m2m3} are obtained by using the normal distribution  of parameters in Table \ref{tab1}.
 The standard deviations are taken to   be 2$\sigma$ for the listed parameter ranges.}
}
 The observed mass ratios  are  successfully reproduced  by taking 
 $k_e \simeq 0.29 -3.0$.
In practice, we take $k_e =3$  in the following numerical calculations.
  It is remarkable that both ratios  $m_e/m_\tau$ and $m_\mu/m_\tau$
  are in a good agreement with observed ones by fixing  one free parameter $k_e$.\footnote{We have assumed $m_\tau \simeq m_b$ at a very high energy scale. However, the mass ratios are almost independent of the radiative corrections.}
 
We see that our model can explain the observation  very well and we will fix it in the following analysis. It should be noted here that the mixing between the second and third families of left-handed charged leptons is already very large because
(2,\,3) and (3,\,3) elements are of the same order 
 as seen in  Eq.\,\eqref{Me}. 
   This large mixing in the charged lepton sector is very important to reproduce the  large $\theta_{23}$ \cite{Esteban:2020cvm} in the  PMNS mixing matrix as pointed out in \cite{Ellis:2002eh} long time ago.

We are now at the point to discuss the neutrino mass matrix. We introduce three heavy Majorana neutrinos $N_{i}~(i$=1-3) to generate not only the observed small neutrino masses via the seesaw mechanism \cite{Yanagida:1979as,Yanagida:1979gs,Gell-Mann:1979vob} \footnote{The right-handed neutrinos are introduced to enhance the $\mu \rightarrow e + \gamma$ decay rate in \cite{Minkowski:1977sc}, where the seesaw mass formula is derived.} but also the Universe's baryon-number asymmetry by the decay of $N_{i}$ \cite{Fukugita:1986hr}. We place them at each fixed points, separately, as like the
$\bf 10_i$ and assign the $\mathbb{Z}_2$ charges for them as the same as those of 
$\bf 10_i$.
See the Table \ref{list} for their charges together with the other fermions
and their locations in the extra two dimensions. The texture of the Dirac mass matrix for neutrinos, $M_D$, has the same structure of the transposed matrix of $M_d$, but values of each matrix elements might be different from those in $M_d^{(1)}$. The Dirac neutrino matrix $M_D$ has six free real parameters, since the CP-violating phase is fixed at  $\phi$ determined from the CKM matrix:
\begin{align}
 M_D=
\begin{pmatrix}
0 & a'_\nu & 0 \\
a_\nu & b_\nu \ {e^{i\phi}}& c'_\nu\\
0 & c_\nu & d_\nu
\end{pmatrix}\,,
\label{MD}
\end{align}
where the Dirac mass term is defined as  ${\cal L}=\bar{\ell}_L M_D N_R$.
The neutrino mass matrix is given by the seesaw mass formula $M_\nu \simeq M_D (1/M_N) M_D^t$, where $M_N$ is the diagonal right-handed Majorana mass matrix, and hence we have additional unknown parameters, that is, $M_1, M_2, M_3$. However, we absorb the $1/\sqrt{M_i}$ in the Dirac mass matrix $M_D$ and hence the number of free parameters in the neutrino mass matrix $M_\nu$ is again six. The redefined matrix $\tilde M_D$ is written by
\begin{align}
\tilde M_D=
\begin{pmatrix}
0 & A' & 0 \\
A & B \ {e^{i\phi}} &C'\\
0 & C & D
\end{pmatrix}\, . 
\label{reMD}
\end{align}
The parameters  $A$, $A'$, $B$, etc.  are defined
by $A= a_\nu/\sqrt{M_1}$, $A'= a'_\nu/\sqrt{M_2}$, $B= b_\nu/\sqrt{M_2}$, etc.
Then, the neutrino mass matrix $M_\nu$ is given as:
\begin{align}
M_\nu=
\begin{pmatrix}
A'^2 & A'B  \,{e^{i\phi}}& A' C \\
 A'B  \ {e^{i\phi}}&  A^2+ C'^2+B^2\, {e^{2i\phi} } &B C \,{e^{i\phi}}+C' D\\
  A' C & B C \,{e^{i\phi}}+C' D\ & D^2+C^2
\end{pmatrix}\,.
\label{Mnu}
\end{align}

It is very interesting that all six free parameters, $A,A',B,C,C'$ and $D$ are determined from the five measured parameters, that is, three mixing angles in the PMNS matrix and
the squared mass  differences as shown in Table \ref{DataNufit}.  We take, in our analysis, the normal  hierarchy (NH) of neutrino masses, that is,  $m_3>m_2>m_1$.
We take $2\sigma$ intervals of the data parameters in Table \ref{DataNufit} to derive our model parameters in Eq.\,\eqref{reMD}. 
\begin{table}[H]
	\begin{center}
		\begin{tabular}{|c|c|c|}
			\hline
			      \rule[14pt]{0pt}{2pt}
\ observable\       &                 best fit\,$ \pm 1\,\sigma$ for NH                  &          2 $\sigma $  range for NH  \\ \hline
			          \rule[14pt]{0pt}{2pt}				                                          $\sin^2\theta_{12}$               &                     $0.308^{+0.012}_{-0.011}$                     &       $0.28-0.33$                \\                 \rule[14pt]{0pt}{2pt}
			$\sin^2\theta_{23}$   &                     $0.470^{+0.017}_{-0.013}$                     &               $0.42-0.59$                \\
			          \rule[14pt]{0pt}{2pt}
			              $\sin^2\theta_{13}$               &                  $0.02215^{+0.00056}_{-0.00058}$            
			              &    $0.021-0.023$    \\ 
			          \rule[14pt]{0pt}{2pt}
$\Delta m_{21}^2$  & $7.49^{+0.19}_{-0.19}\times 10^{-5}{\rm eV}^2$  &     $(7.11-7.87)\times  10^{-5}\,{\rm eV}^2$     \\ 
\rule[14pt]{0pt}{2pt}
$\Delta m_{31}^2$ & \ \ \ \ $2.513^{+0.021}_{-0.019}\times 10^{-3}{\rm eV}^2$ \ \ \ \ & $(2.47-2.56)\times 10^{-3}\,{\rm eV}^2$ \ \ \\  \hline
		\end{tabular}
\caption{The best fit\,$ \pm 1\,\sigma$ and $2\sigma$ range of neutrino  parameters from NuFIT 6.0 for NH (with SK atmospheric data)
		\cite{Esteban:2020cvm}.
		}
		\label{DataNufit}
	\end{center}
\end{table}

 In addition, one should take into account the constraints from the CP Dirac phase
 $\delta_{CP}$ of NuFIT 6.0  \cite{Esteban:2020cvm} and 
 the cosmological bound of the total sum of  neutrino masses \cite{Vagnozzi:2017ovm,Planck:2018vyg,Gerbino:2022nvz}.
{The CP  Dirac phase is given by at the $1\sigma$ level:
\begin{equation}
  \delta_{CP} = (\,212^{+26}_{-41}\,)^{\, \circ}\,,
  \label{CP}
\end{equation}
}
and  the total sum of neutrino masses is given as:
\begin{equation}
  \sum_{i=1}^3 m_i<120\ {\rm meV}\,.
   \label{sum}
\end{equation}
{The constraint Eq.\,\eqref{sum} plays a crucial role for the determination of the six model parameters in the mass matrix $\tilde M_D$. }

We show, in the next section \ref{results}, the predictions of  three physical parameters,
 $\delta_{CP}$, the total sum of  neutrino masses and the effective mass 
 in  the neutrino-less double beta decays $m_{\beta\beta}$.
Those predictions will be tested in near future experiments.

\section{Predictions of low-energy neutrino observables}
\label{results}
Parameters of the  neutrino mass matrix in Eq.\,\eqref{Mnu} are determined  by 
using the neutrino oscillation data in Table \ref{DataNufit} 
with constraints of Eqs.\,\eqref{CP} and \eqref{sum}
since the charged lepton mass matrix in Eq.\,\eqref{Me} is fixed.
(In practice, the charged lepton masses are reproduced within $2\%$ error-bar at the electroweak scale.)
We  obtain the allowed ranges of parameters   by scanning parameters in the real space with
 $|D| \geq |A|, \, |A'|, \,|B|, \, |C|, \, |C'|$ in order to realize  NH of neutrino masses.
 {In practice, we scan the ratios $A/D$,  $A'/D$, $B/D$, $C/D$ and $C'/D$ in the range 
$\pm [0,1]$  with equal weights,
that is the random scan in the linear space.
The absolute value of $D$ is fixed to reproduce the experimental data of $\Delta m^2_{31}$.} 
 The obtained numerical values is listed in Table \ref{tab4}.
 It is noticed that the absolute  values of $A,\,A',\,B,\,C$ 
 are the same order to reproduce the large mixing angle between
  the first and second families, $\theta_{12}$,
   and $\Delta m^2_{21}$.
   On the other hand, $|C'|$ is smaller than other parameters 
   because the large mixing angle between the second  and third  families, $\theta_{23}$,
   is mostly furnished 
   in the charged lepton mass matrix.\footnote{In general, both large mixing angles between the second  and third  families 
   in charged lepton mass matrix  and the neutrino mass matrix
    can reproduce  the observed  $\sin\theta_{23}$ by tuning  some phases.
    However, it is impossible in our scheme  since  we have only one fixed  phase $\phi$. }

\begin{table}[hbtp]
\begin{center}
\begin{tabular}{|c|c|c|c|c|}
\hline 
	\rule[14pt]{0pt}{3pt}  
 $|A/D|$& $|A'/D|$ & $|B/D|$ &  $|C/D|$& $|C'/D|$\\
\hline 
	\rule[14pt]{0pt}{3pt}  
 $0.46-0.49$ & $0.44-0.48$ & $0.09-0.16$& $0.31-0.35$ & $0-0.11$\\
\hline
\end{tabular}
\caption{The allowed regions of parameters in  $M_\nu$.}
\label{tab4}
\end{center}
\end{table}

Because of the precise determination of the model   parameters in Table \ref{tab4}, we can predict 
the Dirac  CP  phase $\delta_{CP}$, the total  sum of  neutrino masses 
and the effective mass $m_{\beta\beta}$ in the  neutrino-less double beta decays of nucleus.

\begin{figure}[h!]
\begin{minipage}[]{0.47\linewidth}
\includegraphics[width=8cm]{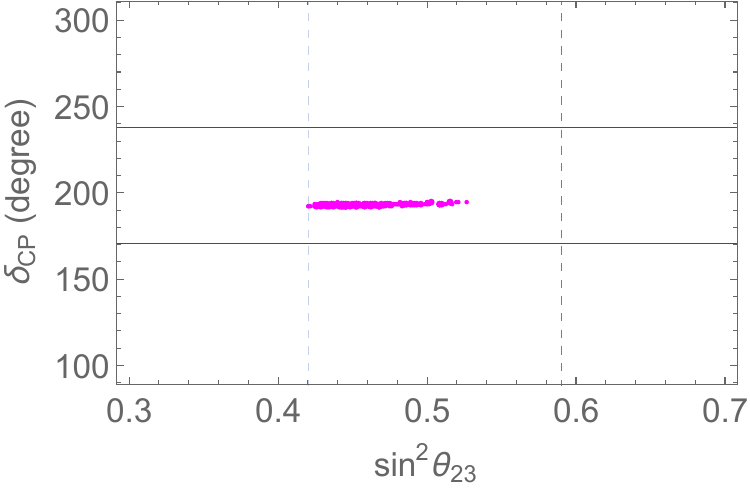}
\caption{The predicted $\delta_{CP}$ versus $\sin^2\theta_{23}$.
The region between the  horizontal red lines denotes $1\sigma$ allowed one of $\delta_{CP}$ in NuFIT 6.0  (NH with SK atmospheric data) \cite{Esteban:2020cvm}. 
The region between blue vertical dashed-lines
denotes 2$\sigma$ allowed one of $\sin^2\theta_{23}$}
\label{fig:CP}
\end{minipage}
\hspace{8mm}
\begin{minipage}[]{0.44\linewidth}
\vskip 0.3 cm
\includegraphics[width=8cm]{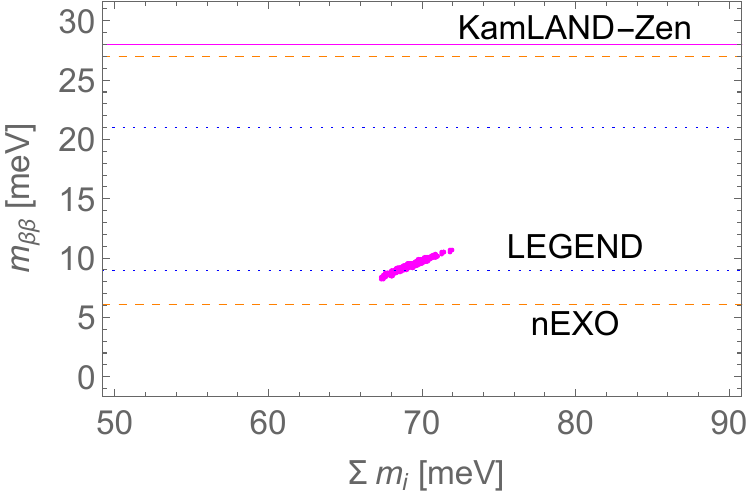}
\caption{The predicted  $m_{\beta\beta}$ versus  $\sum  m_i$.
The magenta solid-line denotes  the upper-bound of KamLAND-Zen 800 \cite{KamLAND-Zen:2024em}
while the blue dotted-line and orange dashed-line denote the sensitivity of 
LEGEND-1000  \cite{LEGEND:2021bnm} and  nEXO \cite{nEXO:2021ujk},
respectively.
}
\label{fig:mass}
\end{minipage}
\end{figure}

At first, we present our prediction of the Dirac CP  phase $\delta_{CP}$ 
versus $\sin^2\theta_{23}$ in Fig.\ref{fig:CP}.
{The predicted range  is $\delta_{CP}\simeq 192^\circ$-$195^\circ$,
 which is inside of $1\sigma$ range of NuFIT 6.0
 (NH with SK atmospheric data) } \cite{Esteban:2020cvm}.
 The Hyper-Kamiokande may detect the signals in  $3\sigma$ level in the future \cite{Hyper-KamiokandeWorkingGroup:2014czz}.
 On the other hand, $\sin^2\theta_{23}$ is not allowed in the full range
 of  $2\sigma$  in NuFIT 6.0 \cite{Esteban:2020cvm}, but   $\sin^2\theta_{23}\simeq 0.42-0.53$.
 This result will be tested in the near future.

 In Fig.\ref{fig:mass}, we present the prediction for
the  effective mass  $m_{\beta\beta}$ in the neutrino-less double beta decays
versus  the total sum of  neutrino masses  $\sum  m_i$.
It is noted  that $m_{\beta\beta}$ is almost proportional to  $\sum  m_i$.
The predicted range of $\sum  m_i$ is $67-73$\,meV,
 which is considerably smaller than the cosmological bound $120$\,meV in Eq.\,\eqref{sum}.
We also predict   $m_{\beta\beta}\simeq 8-11$\,meV.
 Recently, KamLAND-Zen 800 reported a lower limit $28-122$\,meV
\cite{KamLAND-Zen:2024em}
depending on the nuclear matrix elements, which do not yet reach our predicted region,
 while LEGEND-1000  \cite{LEGEND:2021bnm} and 
 nEXO \cite{nEXO:2021ujk} have sensitivity of $9.1-21$\,meV and $6.1-27$\,meV,
 respectively.  Thus, we can expect to test our prediction  in the future.

\section{The baryon-number asymmetry in the present Universe}
\label{Baryon}

The decays of the heavy right-handed neutrinos (RHNs), $N_{i}$, generate the lepton asymmetry which is conversed to the baryon-number asymmetry in the Universe\cite{Fukugita:1986hr}. The dominant asymmetry is generated by the decay of the lightest RHN and thus we have to identify the lightest  RHN to calculate the baryon asymmetry. However, it is very difficult, since we do not have any direct information about the mass hierarchy for the heavy RHNs. In this section, we show that the CP phase relevant to the generation of the baryon asymmetry is directly related to the low-energy CP phase in the CKM and PMNS mass matrices.

Let us assume the $N_i$ is the lightest RHN. The induced lepton asymmetry $\epsilon_i$ is estimated as \cite{Plumacher:1996kc,Hamaguchi:2002vc}
\begin{equation}
    \epsilon_i
    =\frac{\Gamma(N_i\to L H)-\Gamma(N_i\to \bar L \bar H)}
    {\Gamma(N_i\to L H)+\Gamma(N_i\to \bar L \bar H)}\simeq -\frac{1}{8\pi} \sum _{j\not= 1}^3
    \frac{{\rm Im}[(Y_D^\dagger Y_D)^2_{ij}]}{(Y_D^\dagger Y_D)_{11}}
    F^{V+S}\left(\frac{M_j^2}{M_i^2}\right )\,,
    \label{asymmetry}
\end{equation}
where  $Y_D=M_D/v$, $v$ being the vacuum expectation value of the standard Higgs, and
\begin{equation}
    F^{V+S}(x)=\sqrt{x}\left [(x+1)\ln\left (\frac{x+1}{x}\right )-1+\frac{1}{x-1}\right ]
     \xrightarrow[x\rightarrow\infty]{} \frac32 \frac{1}{\sqrt{x}}\,.
\end{equation}
The baryon-number asymmetry is given by
\begin{equation}
    Y_B \simeq -\frac{28}{79}\, \kappa \,\frac{\epsilon_i}{g^*}\,,
\end{equation}
where $\kappa$ is a dilution factor which involves the integration of the full set of Boltzmann equations,   and $g*=106.75$ is taken  in SM.

Recall that only the (2,\,2) element of $M_D$ has the complex phase $\phi$. Thus, the $i$ or $j$ must be 2 to have the non-vanishing complex phase of $(Y_D^\dagger Y_D)^2_{ij}$ 
in Eq.\eqref{asymmetry}. If $i=1$ for instance, we can write down 
$ {\rm Im}[(Y_D^\dagger Y_D)^2_{12}] $  in terms of parameters
of $M_D$ of Eq.\,\eqref{MD} explicitly as:
\begin{equation}
   {\rm Im}[(Y_D^\dagger Y_D)^2_{12}] = {\frac{1}{v^4}\, a_\nu^2\, b_\nu^2\,\sin 2\phi }\,,
\end{equation}
which is {positive} because of $\phi=37^{\circ}-48^{\circ}$ as seen in Table \ref{tab1} and hence the $\epsilon_1$ is {negative}. 
Now, we see that our model predicts the {right} sign of the baryon-number asymmetry if the $N_1$ is the lightest RHN.\footnote{We do not discuss, in this paper, the  value of the asymmetry, since it depends on the details of the early universe such as the reheating processes and temperature and the correct mass of the lightest RHN.}
Important is here that the CP-violating phase for creating the baryon asymmetry is exactly the same as the CP-violating phase $\phi$ at low energies. Thus, it is extremely interesting that the presence of the baryon asymmetry in the Universe predicts  the CP violation observed in the K-meson decay.

\section{Conclusions}

A simple model to solve the strong CP problem without introducing the axion has been recently proposed \cite{Liang:2024wbb}.  We consider, in this paper, an extension of the model to include leptons and the right-handed neutrinos (RHNs) $N_{i}~(i$=1-3). The remarkable point in this extension is that we have only one CP-violating phase $\phi$ and hence all CP-violation parameters in high-energy and low-energy processes are interconnected for each others.  However, the sign of the cosmological baryon asymmetry depends also on mass order of the RHNs and we can not predict the sign of the cosmological baryon asymmetry as long as we do not know the mass hierarchy, as shown in section \ref{Baryon}. We can, however, emphasis that the presence of the baryon asymmetry predicts strongly the CP violation in low energy processes.


On the contrary, we succeed to predict the CP-violating phase in the neutrino oscillation very precisely using the phase $\phi$ determined from the CKM matrix. We obtain
{
\begin{equation*}
   \delta_{CP}\simeq 192^\circ - 195^\circ \,.
\end{equation*}
}
Furthermore, we also predict the important mass parameter $m_{\beta\beta}$  as 
\begin{equation*}
    m_{\beta\beta} \simeq 8-11\  [{\rm meV}]\,,
\end{equation*}
which is the key parameter for the neutrino-less double beta decay. Both predictions will be tested in near future experiments.

The present model is based on only one mass matrix, $M_d^{(1)}$, in \cite{Tanimoto:2016rqy}, but we can construct variant models taking other consistent mass matrices described in  \cite{Tanimoto:2016rqy}. It is a straightforward application of our model construction which will be done in future publications.


\section*{Acknowledgments}
T.~T.~Y.~thanks Qiuyue Liang for discussion about two-zeros texture of quark and lepton mass matrices. This work was in part supported by MEXT KAKENHI Grants No.~24H02244 (T.~T.~Y.). T.~T.~Y.~was supported also by the Natural Science Foundation of China (NSFC) under Grant No.~12175134 as well as by World Premier International Research Center Initiative (WPI Initiative), MEXT, Japan. 



\bibliography{ref}{}
\bibliographystyle{JHEP}

\end{document}